\documentstyle[aaspp4,flushrt]{article}

\def\plotone#1{\centering \leavevmode
\epsfxsize=\columnwidth \hfil \epsfbox{#1}}

\begin{document}

\twocolumn
\title{Energetic Young Radio Pulsars}
\author{J. F. Bell\altaffilmark{1}}
\altaffiltext{1}{The University of Manchester, NRAL, Jodrell Bank,
Cheshire SK11~9DL, UK.~~~ Email: jb@jb.man.ac.uk}

\begin{abstract}
Young radio pulsars shed vast amounts of rotational energy, sometimes as
high as 100,000 times the total energy loss rate from the sun. The wide
range of phenomena resulting from this energy loss include: glitches,
timing noise, jets, bow shocks, bullets and plerions and are be reviewed from
an observational perspective. Past and proposed surveys for young radio
pulsars are summarised along with pulsar birth velocities and
associations with supernova remnants. There are now 4 radio pulsars with
measured braking indices. The resulting constraints on the evolution of
young radio pulsars are discussed in light of the presently observed
population of pulsars. Observations at optical, X-ray and gamma ray
energies which provide a unique opportunity to study the emission and
magnetospheric processes are described briefly. The status of pulsar
birth velocities and supernova remnant associations are summarised.
\end{abstract}

\section{Introduction}

This review summarises the wide range of phenomena that occur in, or
are associated with, energetic young radio pulsars. The scope is
confined to rotation powered radio pulsars as accretion powered pulsars are
discussed by others at this meeting. The level is introductory,
providing the background for other speakers who discuss particular
aspects in more detail. From here on rotation powered radio pulsars are
referred to as pulsars and energetic young rotation powered radio pulsars
are referred to as young pulsars. As a working definition we 
consider a pulsar with a characteristic age less than 100,000 years to be
young. A key point to be emphasised is that many phenomena associated with
young pulsars arise from the very high spin down energies that they deposit
in their surroundings.

\noindent
The layout of this review is as follows:

\noindent
$\S$\ref{s:prop} Basic properties of young pulsars\\
$\S$\ref{s:surv} Surveys for young radio pulsars\\
$\S$\ref{s:gli}  Glitches, timing noise and braking indices \\
$\S$\ref{s:hen}  High energy observations\\
$\S$\ref{s:ism}  Interstellar medium interactions \\
$\S$\ref{s:vel}  Birth velocities\\
$\S$\ref{s:ass}  Associations with supernova remnants

\section{Basic Properties of Young Pulsars}
\label{s:prop}

\begin{figure}
\plotone{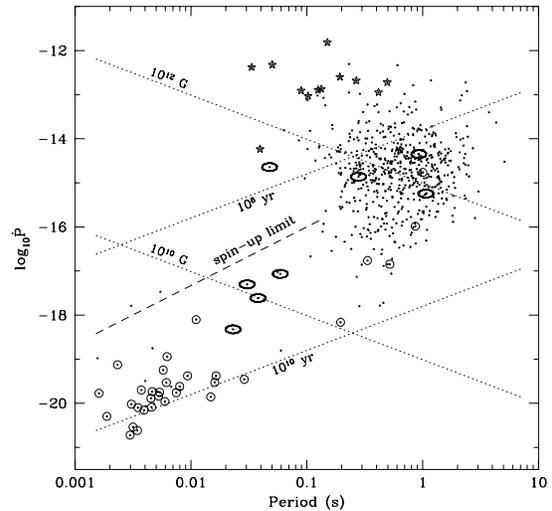}
\caption{$P$-$\dot{P}$ diagram for radio pulsars. Circles indicate pulsars
in circular orbits, ellipses indicate pulsars in eccentric orbits and stars
indicate pulsars associated with supernova remnants.}
\label{f:ppd}
\end{figure}

The age of a pulsar with period $P$ and period derivative $\dot{P}$, is
given by
\begin{equation}
\label{e:age}
t = {P \over (n-1) \dot{P}} \left( 1 - \left( {P_{0} \over P} \right)^{(n-1)}
\right)
\end{equation}
where $P_{0}$ is the rotation period at birth and $n$ is the braking
index. $P_{0}$ can only be determined in a couple of special cases and it is
generally assumed that $P_{0} \ll P$. It is further assumed that magnetic
dipole braking is the dominant braking force, for which $n = 3$. Applying
these assumptions to (\ref{e:age}) gives the characteristic age $t_{c} =
P/2\dot{P}$. Lines of constant characteristic age are shown in Figure
\ref{f:ppd} and suggest that most pulsars are $10^{6} - 10^{7}$ years
old. In the middle at the top of Figure \ref{f:ppd} many of the young
pulsars are shown with starred symbols which indicate that they are possibly
associated with supernova remnants (SNRs).

\begin{figure}
\plotone{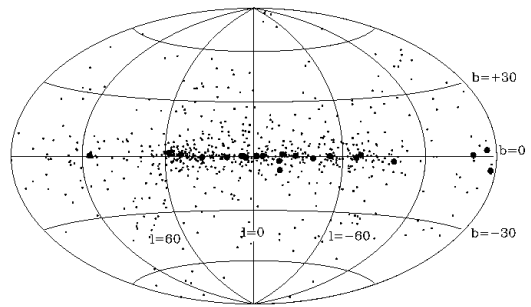}
\caption{An Aitoff projection of the Galactic pulsar population. Young
pulsars are shown with large dots.}
\label{f:gpy}
\end{figure}

\begin{figure}
\plotone{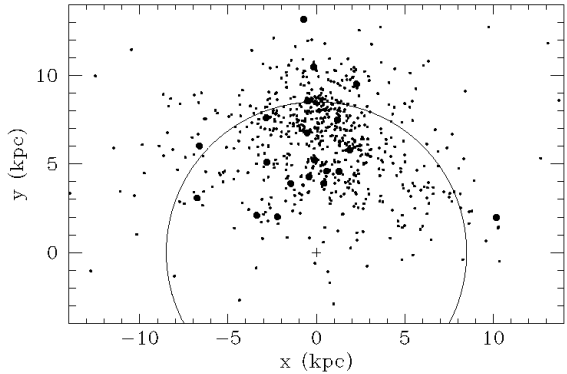}
\caption{A face-on view of the Galactic pulsar population. Young
pulsars are shown with large dots.}
\label{f:gfy}
\end{figure}

\begin{figure}
\plotone{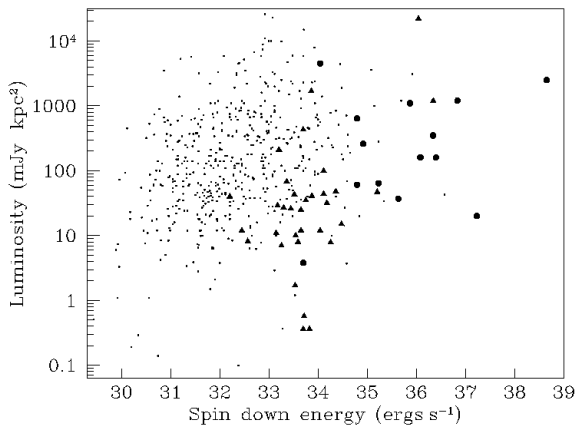}
\caption{The population of pulsars with measured radio luminosities and spin down
energies. Young pulsars are shown with large dots and MSPs with triangles}
\label{f:ledot}
\end{figure}

The median values of a range of pulsar parameters are compared in Table
\ref{ypar} for three populations of pulsars: young, normal and millisecond
pulsars (MSPs). The most noticeable unique features of young pulsars are
the large spin down energies ($\dot{E}$) and their small distance ($z$)
from the galactic plane. This very low height above the galactic plane is
striking (Figure \ref{f:gpy}). They also appear to have slightly larger
magnetic fields ($B$), dispersion measures (DM) (and therefore distances,
as shown in Figure \ref{f:gfy}) and Luminosities ($L$). However as can be
seen in Figure \ref{f:ledot}, the scatter in the luminosities is large for
the young pulsars, while the MSPs are more tightly clustered. The high spin
down energies of both these population relative to the main population is
also very clear in Figure \ref{f:ledot}. Rather surprisingly the observed
young population is now smaller than the observed MSP population.

The group of presently known young pulsars ($t < 100,000$ years) are listed
in Table \ref{yng} along with several of their important properties which
are summarised in column eight. There are three very young pulsars with ages
just over 1000 years, PSRs B0531+21 (Crab) and B1509--58 in the Galaxy and PSR
B0540--69 which is one of the four known pulsars in the large Magellanic
cloud. PSR J0633+1746 (Geminga) is included in the list due to its
similarity to radio pulsars, even though it has not been detected at
radio wavelengths.

\section{Surveys for Young Radio Pulsars}
\label{s:surv}

\subsection{Selection Effects}

Most pulsars are born in or near the Galactic plane where the progenitor
population is located. Since most young pulsars will not have had time to
move very far from the Galactic plane they too are found most abundantly in
the Galactic plane (Figure \ref{f:gpy}). This has important implications for
surveys for young pulsars, including some severe selection effects. We begin
by discussing those which broaden the pulse profile. The importance of this
can be seen from the fact that the sensitivity of a pulsar survey depends on
the square root of the observed duty cycle of the pulsars it seeks to find
(\cite{jlm+92}). That is, a large duty cycle leads to a poorer sensitivity.

Dispersion smearing across individual frequency channels (Bhattacharya,
these proceedings) can be significant. For example, in the recent Parkes
survey (\cite{mld+96,lml+97}) the dispersion smearing of a pulsar with a
dispersion measure of $DM=1000$ cm$^{-3}$\,pc would be 12.8ms. This smearing
is strongly frequency dependent and is given by
\begin{equation}
\label{e:dms}
\Delta t = \left( {204 \over f} \right)^{3} DM~~~ms/MHz
\end{equation}
where $f$ is the radio frequency at which the observations are made.  If a
frequency a factor of 3 higher is used, then dispersion smearing is reduced
by a factor of 27 and the sensitivity potentially improved by a factor of
5!  This problem could also be overcome by using much narrower frequency
channels, but the smearing caused by interstellar scattering cannot
(Bhattacharya, these proceedings). Here the pulse broadening depends on
$f^{-4}$, again suggesting that typical survey frequencies of $\sim
400$~MHz are not optimal for finding young pulsars. As young pulsars are
relatively luminous, they may be found at large distances if the increased
dispersion smearing and scattering can be countered by observing at high
frequencies.

The frequency dependence of the sky background and the spectral indices of
pulsars are also important to consider when designing a survey. The sky
background temperature close to the Galactic plane can be as high as 400\,K
at 400\,MHz, completely dominating the receiver noise. However, between 400
and 1500\,MHz the sky background has a frequency dependence of $f^{-2.6}$,
resulting in a background temperature of $\sim 20$\,K at 1400 MHz,
comparable to a well optimised receiver. The flux detected from young
pulsars typically depends on $f^{-2}$ for frequencies between 400 and
1500\,MHz. While this means that the observed fluxes are smaller at higher
frequencies, the dependence on frequency is flatter than the dependence for
dispersion smearing, scattering and the sky background.

\subsection{Past, Present and Future Surveys}

A moderately high frequency search provides an optimal approach for finding
young pulsars. Such searches have already been undertaken
(\cite{jlm+92,clj+92}) and were very successful.  They found a sample of 86
pulsars in a region of sky that had already been well searched at lower
frequencies. The mean age of this sample is 10\% of the mean age of the
entire pulsar population. These two surveys are summarised together with
three next generation high frequency surveys for young pulsars in Table
\ref{surv}. The Nancay survey is a collaboration between Meudon, Berkeley
and Naval Research Labs. and will provide excellent sensitivity to young
pulsars, with vastly improved time and frequency resolution and limiting
sensitivity.

One of the difficulties of high frequency surveys is the slower sky coverage
due to the smaller beam. For example a 1400\,MHz survey would take 12 times
as long as a 400\,MHz survey to cover a given region with the same
integration time and would collect substantially more data. At Parkes and
Jodrell Bank, a novel approach of using many simultaneous beams is being
used to speed up the sky coverage. Together with better receivers, this will
improve the sensitivity by a factor of 5-10 for the same amount of telescope
time per square degree as the Clifton et al (1992) and Johnston et al (1992)
surveys. The corresponding longer integration time on each pointing means
that acceleration searches will be necessary in order to find pulsars in
fast binaries.

\section{Glitches, Timing Noise and Braking Indices}
\label{s:gli}

\subsection{Glitches}

In general, radio pulsars are excellent clocks, particularly the MSPs. There
are however many pulsars which show substantial departures from a regular
slow down, making the rotation unpredictable. These departures are manifest
in two forms, sudden instantaneous spin ups called glitches and random walk
wanderings called timing noise. Both these effects are most prevalent in
young pulsars, but do also occur in older pulsars. The unpredictability and
randomness of these effects leads to considerable difficulties in
determining braking indices. On the other hand they result from the
dynamics of the interior of neutron stars and thereby provide a window
through we can hope to peer into the interior.

Alpar (these proceedings) has discussed the nature and theories of glitches
in detail. The basic properties of glitches include a step in rotation
period of $\Delta P/P = 10^{-9} - 10^{-5}$, a step in rotation period
derivative $\Delta \dot{P}/\dot{P} \sim 10^{-3}$ and an exponential recovery
(\cite{lss97}). The similarity of $\Delta \dot{P}/\dot{P} $ for most
glitches provides indisputable evidence both for the vortex pinning model
and evidence against the crust cracking model (Alpar, these proceedings).
Long term regular monitoring of old favourites including the Crab and Vela
has been continued and several new glitches have been observed
(\cite{lps93,arz95}) including a giant glitch in PSR~B1757$-$24
(\cite{lkb+96}). Some groups continue to monitor individual pulsars such as
Vela for up to 18 hours per day (\cite{mcc96}).

At the time of writing, 52 glitches have been observed in 21 pulsars. These
21 pulsars have a median age of 150,000 years, with 10 of them having ages
less than the (arbitrary) 100,000 year boundary between young and other
pulsars. Of the 52 observed glitches, 40 have occurred in young pulsars,
showing that young pulsars glitch more often than older pulsars. There have
already been 21 glitches detected in the pulsars found in the high frequency
surveys (Section \ref{s:surv}). This large number of glitches has now
allowed statistical studies which indicate that post-glitch relaxations can
be separated into two exponential components for most pulsars (\cite{sls96}).

Glitches are associated with high $\dot{P}$ young pulsars and are rare. Most
of the pulsars that have glitched have done so only once, indicating that
the time scale for glitches is long compared to the observation time scale
of a few years.  Many of the pulsars near the top right of this plot have a
negative $\ddot{P}$ which is an indication that these pulsars have glitched
in the past (\cite{lyn96}). From an analysis of many glitches Lyne (1996)
\nocite{lyn96} also demonstrated that glitch activity peaks for pulsars with
ages of $\sim 20,000$ years.

\subsection{Timing Noise}

Timing noise is characterised by random walk phase wandering of the pulses
relative to a simple slow-down model.  Timing noise is generally very red
and the timing residuals relative to a simple slow-down model are often
dominated by a cubic term. Hence, the second time derivative of the period
$\ddot{P}$ provides a simple way (there are many others) of measuring the
amount of timing noise. Timing noise is greatest in young pulsars with large
$\dot{P}$ (\cite{lyn96,antt94}). While most millisecond pulsars have
very small period derivatives and are found to be very stable, PSR B1937+21,
which has a relatively high $\dot{P}$ for a millisecond pulsar, may have
shown detectable timing noise (\cite{ktr94}).

An enormous timing database (equivalent to over 3500 years of data on a
single pulsar) has been collected at Jodrell Bank and is presently being
analysed (\cite{mlp97}). This will allow a very complete understanding of
timing noise and improve on previous studies (\cite{arz95}). The time span
of data collected for a number of MSPs will soon be long enough to assess
the extent to which timing noise occurs more generally in the fastest
pulsars (\cite{tay96}). In this respect the next few years should establish
whether or not pulsars can really be used as long term standards of time.

\subsection{Braking Indices and Spin down evolution}

The second derivative of the period can be measured for many pulsars. The
vast majority however, are corrupted by timing noise and are
therefore not associated with steady spin down. As a result there are only
four pulsars for which the measured braking indices ($n$) are believed to be
associated with the steady spin down (\cite{lpgc96}):\\

\begin{tabular}{ll}
$\bullet$ B0531+21  & $2.51 \pm 0.01$  \\
$\bullet$ B0833--45 & $1.4  \pm 0.2$   \\
$\bullet$ B1509--58 & $2.837\pm 0.001$ \\
$\bullet$ B0540--69 & $2.24 \pm 0.04$  \\
\end{tabular}\\

If braking indices are so low for most pulsars (recall $n=3$ for magnetic
dipole braking), this not only has implication for the braking mechanism,
but also for the implied velocities of pulsars claimed to be associated with
SNRs. The true ages of pulsars would be greater and hence the implied
velocities would be smaller (\cite{lpgc96}). If the evolutionary tracks based
on the above values for $n$ are compared with the present position of these
pulsars and the position of the main population in Figure \ref{f:ppd}, it
becomes clear that these 4 pulsars are not evolving towards the main
population (\cite{cam96b}).  There are several question that arise from
this. Why is the area of the $P - \dot{P}$ diagram to which these pulsars
are evolving devoid of pulsars? Where are the young pulsars that are feeding
into the main population? Will $n$ increase later in the evolution so that
these pulsars do enter the main population? Even though there are more
theories and selection effects than there is room to discuss here, the
answer to the above puzzle is not clear.
 
\section{High Energy Observations}
\label{s:hen}

Radio timing observation of pulsars provides precision of measurement that is
unlikely to be surpassed by other techniques. This precision has led to an
astounding range of experiments and measurable effects that have probed
fundamental physics at the deepest level. However, when it comes to studying
the emission mechanisms and magnetospheres of pulsars the story is
different. Such a small fraction ($< 0.1$\,\%) of the spin down energy is
emitted at radio wavelengths that we often hear the analogy: ``Trying
to understand the emission mechanisms and magnetospheres of pulsars is like
standing outside a factory with no windows and trying to work out what they
are making inside by listening to the noises from the factory''.

In the X-ray and $\gamma$-ray bands the fraction of the spin down energy
radiated approaches a few percent. This then gives a much better constraint
on the emission and magnetospheric processes and is one of the prime purpose
for which these high energy observations are made. The relevance to young
pulsars is straight forward. All six of the pulsars detected in
$\gamma$-rays are young (Kanbach, these proceedings). A large fraction of
the pulsars detected in X-rays are young (\cite{bt97}). Young pulsars
detected at higher energies are noted in column 8 of Table \ref{yng}.

Thermal components from the cooling of the neutron star surface are detected
for 3 pulsars of which only Geminga is young. These observations provide
direct constraints on the nature and composition of the neutron star
atmospheres. However, when the thermal components are removed and the
non-thermal components are considered for all the X-ray detected pulsars
there is a strong linear correlation between the X-ray luminosity and the
spin down energy (\cite{bt97}). This correlation extents over several orders of
magnitude and is remarkable given the uncertainties in the distances of at
least 30\% (\cite{tc93}).

\subsection{Timing}
\label{s:htim}

Radio pulsar timing at wavelengths other than radio has until recent times
been mostly restricted to two pulsars, the Crab and PSR~B0540$-$69 at
optical wavelengths, although there was also some timing of Vela
(\cite{mwpe80}). Caraveo and others (this proceedings) give more detailed
discussions of timing at these wavelengths. There was extensive optical
timing of the Crab during the 1980's (\cite{loh81}) which resulted in the
detection of glitches (\cite{loh75,gro75c}) and also studies of the random
walk nature of timing noise (\cite{cor80}). Recently there have been
simultaneous timing studies of the radio giant pulses and gamma ray
pulsations (\cite{lcu+95}). PSR~B0540$-$69 was discovered in X-rays
(\cite{shh84}) but most of the timing including the measurements of the
braking index has been at optical wavelengths )\cite{mpb87,mp89,gfo92}). It
was eventually detected in the radio after some very long integrations by
radio search standards (\cite{mml+93}). Recently braking indices for the
Crab and PSRs B0540$-$69 and B1509$-$58 have been determined from X-ray
timing observations using Ginga (\cite{ndl+90}). The gamma ray pulsar
Geminga has not been detected at radio wavelengths but is mentioned here due
to its similarity to Vela and other radio pulsars. There has been a
considerable amount of timing of Geminga, in particular using COS B
(\cite{bc92,gbb+94}) and SAS 2 (\cite{mbf92}) which accumulated over 10
years of timing data.

\section{ISM Interactions}
\label{s:ism}

\subsection{Jets and/or Plumes}
\label{s:jet}

There are a number of pulsars which may be responsible for jets (plumes),
though none of these jets (plumes) are known to be relativistic.  The X-ray
band offers the most prospects, with claims of X-ray jets for at least 6
pulsars. In some cases (PSRs B0355+54, B1055$-$52, B1929+10) these linear
X-ray sources can be interpreted as mini-Crab nebulae, resulting from
non-collimated relativistic particles confined by ram pressure as the pulsar
moves through the ISM, leaving a wake of X-ray emission
(\cite{fei86,hel84a,yhh94}).

There are several cases in which this interpretation does not fit with the
observations and the case for collimated emission is stronger. For the Vela
pulsar (B0833$-$45) the X-ray jet discovered using the Einstein observatory
(\cite{hgsk85}) does not line up with the pulsar proper motion
(\cite{bmk+89,car96}). Using a ROSAT image Markwardt and 
\"{O}gelman (\cite{mo95}) interpreted the jet as thermal,
while recent ASCA observations seemed to reveal a non-thermal spectrum
(\cite{kaw96}). The nature of the spectrum still seems to be uncertain
(Markwardt, these proceedings. There are strongly polarised linear features
in the radio maps of Milne (1995) \nocite{mil95} that may be associated with
the X-ray jet, although these have not been interpreted as a radio
jet. Tamura et al. (1996) \nocite{tkyb96} recently used ASCA observations of
PSR~B1509$-$52 to suggest that it has both a non-thermal X-ray jet generated
by the pulsar and a thermal X-ray nebula that is created at the working
surface of the jet with the ISM.

Apart from the Crab pulsar (see \S\ref{s:cjet}), there is little evidence
for jets from radio pulsars at other wavelengths, despite some concerted
searches for them. In particular, at radio wavelengths the regions around
numerous pulsars have been mapped using the VLA (\cite{ccgm83}), with no
convincing evidence for jets, except for PSR~B1610$-$50 in the supernova
remnant Kes 32 (\cite{rmk+85}). Maps at 843 MHz show a well collimated
(although thermal, non-polarised) jet emerging from the SNR and then
spreading into a wide plume. Only the main part of the remnant is detected
at X-ray wavelengths (\cite{kaw96}). Many deep optical images have been
obtained in order to study emission from the pulsars or their
companions. The only pulsar (other than the Crab) with a candidate optical
jet is PSR~B0540$-$69 (\cite{car96}). This was obtained with HST and is seen
as a weak double sided linear feature lying across the pulsar in an
H${\alpha}$ light.

\subsection{The Crab Pulsar and its Nebula}
\label{s:cjet}

With no fewer than {\it three} jets the Crab pulsar must be considered to be
as bizarre as the other Galactic jet sources such as SS433 and
GRO~J1655$-$40. The three jets are referred to as southeast, northwest
and north.

A ROSAT image of the inner Crab nebula (\cite{asc92}), confirmed the
presence of the torus (\cite{bal85}) and revealed a bright jet southeast of
the pulsar. This jet appears to be aligned with the spin axis and is
perpendicular to the plane of the torus (\cite{hss+95}) indicating the
presence of collimated relativistic particles from the poles. With
increasing distance from the pulsar, it becomes wider and bends
southward. Wide field ground based images reveal a very similar but slightly
more extended structure in line free optical continuum light. High
resolution HST images in a similar band reveal two knots at 1,400\,AU and
10,000\,AU from the pulsar due to shocks in the inner jet
(\cite{hss+95}). Future images of a similar quality and resolution should
allow the measurement of the proper motion of the knots, giving some
indication of the velocity of the jet. The VLA map of Bietenholz and
Kronberg (1990) \nocite{bk90a} shows no evidence for a corresponding radio
jet.

A similar but less well defined X-ray  structure in the opposite (northwest)
direction is interpreted as material compressed by the counter jet. In the
optical images there are several features which run parallel to the edges of
the counter jet. From a comparison of images taken at different epochs these
have been shown to have proper motions perpendicularly away from the jet
(\cite{hss+95}). Again in the radio maps, there is no evidence for a
radio counter jet (\cite{bk90a}). 

A third unrelated jet in a northerly direction, associated with the Crab
nebula was found in a deep optical III\,aJ plate (\cite{van70}), although it
was also present on isophote drawings (\cite{wol57}). The jet was detected in
the emission lines O{\scriptsize III}, H${\alpha}$ and N{\scriptsize II},
indicating that it has a non-thermal origin (\cite{gf82}). It is clearly
present in radio maps made with the VLA (\cite{vel84,fkcg95}). The jet is
highly polarised and non-thermal with the local magnetic field lines running
parallel to the jet (\cite{wsh85,bk90a}). Recent proper motion measurements by
Fesen and Staker (1993) \nocite{fs93} and Marcelin et al. (1990)
\nocite{mvw+90} are consistent with less precise earlier results and show
that the jet is expanding along its length at 2500 km\,s$^{-1}$ and
perpendicular to its length at 260 km\,$^{-1}$. Using these velocities and
the present jet size suggest that it formed around the same time as the
pulsar. Unlike the other jets, this jet does not appear to be replenished by
the pulsar.

\subsection{Bow Shocks, Bullets and Plerions}
\label{s:bow}

There are a number of objects associated with pulsars which result from the
emission of relativistic particles which are not collimated. These are
briefly mentioned in this paragraph, with references to recent
reviews. Pulsar wind nebulae result from bow shocks where pulsar winds are
balanced by ram pressure in the interstellar medium (ISM)
(\cite{cor96}). While these objects are rare, they have been detected in
H${\alpha}$, soft X-rays and radio and and may be used to determine pulsar
distances, radial velocities, ISM neutral hydrogen content and the soft
X-ray composition of the pulsar winds. Recently many have been observed in
X-rays using ASCA (\cite{kaw96}). There are presently four eclipsing pulsars
known and all of these are millisecond pulsars (\cite{fru96}). In such
systems, a significant fraction of the pulsars' flux at all wavelengths
impinges on the companion and heats it. For two of the four systems this is
clearly observed as an optical brightening of the companion at superior
conjunction, and offers constraints on the composition of the pulsar
wind. Finally, there is the case of bullets around the Vela supernova
remnant (\cite{aet95,sjva95}). These are detected in both X-rays and radio
and appear to be lumps of material ejected at the time of the supernova
explosion.

\section{Birth Velocities}
\label{s:vel}

For many years, there has been a growing body of evidence that radio pulsars
are born with large recoil velocities: \\

\noindent
$\bullet$ Large scale height of pulsars (\cite{lmt85})\\
$\bullet$ Large scale height of LMXBs (\cite{vw95})\\
$\bullet$ Large observed proper motions (\cite{hla93})\\
$\bullet$ Pulsars found outside SNRs (\cite{fgw94})\\
$\bullet$ Bow shocks in the ISM (\cite{cor96})\\

Lyne and Lorimer (1994) \nocite{ll94} demonstrated that the main population
of pulsars is devoid of many high velocity pulsars. This is because during
their observable lifetime, their velocity is large enough to have allowed
them to leave the Galaxy. In order to determine the mean pulsar birth
velocity of $450 \pm 90$ km\,s$^{-1}$ they used a sample of young
pulsars. The present small sample of young pulsars is a fundamental
limitation and the many young pulsars that should be found in the new surveys
(Section \ref{s:surv}) will redress this problem.

\subsection{Evidence for Kicks from Asymmetric Supernovae}

The mean birth velocity of $450 \pm 90$ km\,s$^{-1}$ is 3 times higher than
previous estimates. One of the most important questions associated with this
is whether such velocities can be produced by orbital motions, or whether
the neutron stars receive kicks from asymmetric supernova explosions. Van
den Heuvel and van Paradijs (1997) \nocite{vv97} summarised several pieces
of evidence that cannot be explained without kicks: \\

\noindent
$\bullet$ Misaligned spin and orbit of PSR J0045--7319 (\cite{kbm+96})\\
$\bullet$ Misaligned spin and orbit of PSR B1534+12 (\cite{wol92})\\
$\bullet$ Eccentricities of Be-Xray binaries (\cite{vv95})\\
$\bullet$ Rarity of double neutron star systems (\cite{bai96})\\
$\bullet$ Rarity of Galactic LMXBs (\cite{vv97})\\

All of these are very strong evidence in favour of kicks, except possibly the
misaligned spin and orbit of PSR B1534+12, since the interpretation of the
polarimetry is still ambiguous (\cite{aptw96}). The following additional pieces
of evidence also support the kick hypothesis:\\

\noindent
$\bullet$ No pulsars found in OB and Be star surveys (\cite{pelf96})\\
$\bullet$ Kicks required to produce pulsar spin rates (\cite{phi97})\\
$\bullet$ Misaligned spin and orbit of PSR B1259--63 (Melatos et al 1995)
\nocite{mjm95}\\ 
$\bullet$ Rarity of Galactic MSPs (\cite{tb96})\\

Recently Phinney (1997) \nocite{phi97} reiterated the difficulty of forming
a rapidly rotating neutron star directly from the core collapse during a
supernova explosion. He demonstrated that a misdirected birth kick could
provide both the extra angular momentum required for the rapid spin as well
as the observed recoil velocities. This mechanism suggested that for a given
strength of birth kicks, there would be an anticorrelation between spin
rate and recoil velocity.   Another implication is that a significant
fraction of young pulsars would be born with periods ranging 1 to 1000
ms. We should therefore see some pulsars in supernova remnants with long
periods. While there may be selection effects against finding such pulsar
notable examples are PSRs~B2334+61 ($P=495$\,ms) and B1758--23 ($P=416$\,ms)
(\cite{kf93}).

If Phinney (1997) is correct then the population of truly young pulsars may
include many pulsars with considerably lower spin down energies than those
which are the main subject of this review.

\section{Associations with Supernova Remnants}
\label{s:ass}

The association of neutron stars with supernova remnants (SNRs) began with
Baade and Zwicky (1934) \nocite{bz34}. The discovery of the Crab pulsar in the
Crab nebular provided the first evidence supporting their bold proposition
(\cite{sr68}). The typical observable lifetime of SNRs is comparable to the
ages of the young pulsars we are discussing here, so that many young pulsars
are likely to be found in or near the remnants from the supernovae which
formed them. The task of identifying which pulsars are associated with which
SNRs is difficult. There are now some 215 SNRs known (\cite{gre96}) and the
probability of chance positional coincidences is significant.

Kaspi (1996) \nocite{kas96} listed several additional questions that require
a positive answer for a pulsar and SNR to be physically associated:\\

\noindent
$\bullet$ Do independent distance estimates agree ?\\
$\bullet$ Do independent age estimates agree ?\\
$\bullet$ Is the implied velocity reasonable ?\\
$\bullet$ Is there any evidence for interaction ?\\
$\bullet$ Are the pulsar proper motion and position offset consistent ?\\
$\bullet$ Are the overall statistics sensible ?\\

In many cases, the pulsar is well outside the remnant that it is claimed to
be associated with. As a result there are several cases in which the implied
velocity is over 1000 km\,s$^{-1}$.  Given the uncertainty in the distances
and ages, particularly for SNRs, the most pertinent of the above
questions is the consistency of the pulsar proper motion and its present
position with respect to the remnant. Unfortunately proper motion
measurements are available for only a few of the 28 pulsars which are
claimed to be associated with supernova remnants.

Kaspi (1996) \nocite{kas96} drew up a score sheet for the 28 claimed
associations based on the above questions. A summary of the results for
young pulsars is given in the last column of Table \ref{yng}, with 6
definite associations and 5 probable. Including PSR B1951+31 (age = 107,000
years) gives a total of 7 definite associations. The remaining 16 candidate
associations are either not real or lack sufficient observational evidence
to determine their credibility. Similar questions are addressed by Frail
(these proceedings) who provides a census of all types of neutron stars
associated with SNRs.

\acknowledgements 
I thank Ed van den Heuvel, Duncan Lorimer and Dale Frail for discussions on
pulsar velocities, Jean-Francois Lestrade for unpublished parameters, Andrew
Lyne and Graham Smith for helpful discussions and the organisers for a fun
and productive meeting.


\begin{table}[h]
\begin{center}
\caption{Median values of parameters for three observed populations of
pulsars}
\label{ypar}
\begin{tabular}{l|ccc} \hline 
                 & Young & Normal  & MSP  \\ \hline
$P$ (ms)         & 150   & 622     & 4.6  \\
$\dot{P}$    & $1.2 \times 10^{-13}$ & $1.8 \times 10^{-15}$ & $2.0 \times
10^{-20}$ \\
Age (years)  & $2.0 \times 10^{4}$   & $5.5 \times 10^{6}$   & $5.0 \times
10^{9}$ \\
$B$ (Gauss)    & $4.9 \times 10^{12}$  & $1.0 \times 10^{12}$  & $3.6 \times
10^{8}$ \\
$\dot{E}$ (ergs\, s$^{-1}$) & $1.3 \times 10^{36}$ & $2.4 \times 10^{32}$ &
$5.0 \times 10^{33}$ \\
DM (cm$^{-3}$\,pc) & 219   & 89      & 22   \\
Dist. (kpc)      & 4.3   & 3.8     & 1.3  \\
z (kpc)          & 0.04  & 0.35    & 0.38 \\
S400 (mJy)       & 10    & 11      & 15   \\
L (mJy kpc$^2$)  & 260   & 130     & 35   \\ 
Number           & 23    & 651     & 59   \\ \hline
\end{tabular}
\end{center}
\end{table}

\onecolumn

\tabcolsep 3pt
\begin{table}[p]
\begin{center}
\caption{Young Energetic Radio Pulsars. O -- optical, X -- X-rays, $\gamma$ --
$\gamma$-rays, I -- infra-red, J -- jet, N -- nebula, G -- glitches, B -- bow shock, Y
-- yes, P -- probably, ? -- maybe, U -- unlikely}
\label{yng}
\footnotesize
\begin{tabular}{lcrcccrlc}\hline 
   Name    &    P  &  Pdot & Age & $\log(\dot{E})$ &$\log(B)$ &    b
& Comment        &  SNR \\ 
           & (sec) &  (10$^{-15}$) & (kyr)& (ergs\,s$^{-1}$)    & (Gauss)          &     
&                &     \\ \hline 
B0531+21   & 0.033 &  420.9 & 1.2 & 38.6 & 12.6 &  --5.8 & O,X,$\gamma$,I,J,N,G   & Y \\
B1509--58  & 0.150 & 1536.5 & 1.5 & 37.2 & 13.2 &  --1.2 & X,$\gamma$,J,N         & P \\
B0540--69  & 0.050 &  479.0 & 1.6 & 38.2 & 12.7 & --31.5 & O,X                    & Y \\
B1610--50  & 0.231 &  496.0 & 7.3 & 36.2 & 13.0 &  +0.2  & X,J,N                  & ? \\
B0833--45  & 0.089 &  124.1 & 11  & 36.8 & 12.5 &  --2.8 &
O,X,$\gamma$,J,N,G     & Y \\ \\
B1338--62  & 0.193 &  253.2 & 12  & 36.1 & 12.9 &  --0.0 & G                      & Y \\
B1757--24  & 0.124 &  128.0 & 15  & 36.4 & 12.6 &  --0.9 & G                      & Y \\
B1800--21  & 0.133 &  134.3 & 15  & 36.3 & 12.6 &  +0.2  & X,G                    & P \\
B1706--44  & 0.102 &   93.0 & 17  & 36.5 & 12.5 &  --2.7 & X,$\gamma$,G           & ? \\
B1046--58  & 0.123 &   96.0 & 20  & 36.3 & 12.5 &  +0.6  & X,N
& U \\ \\
B1737--30  & 0.606 &  465.3 & 20  & 34.9 & 13.2 &  +0.2  & G                      & U \\
B1853+01   & 0.267 &  208.4 & 20  & 35.6 & 12.9 &  --0.5 & B                      & Y \\
B1823--13  & 0.101 &   74.9 & 21  & 36.5 & 12.5 &  --0.7 & X,G                    & ? \\
B1727--33  & 0.139 &   85.0 & 25  & 36.1 & 12.5 &  +0.1  & G                      & ? \\
B1643--43  & 0.231 &  112.7 & 32  & 35.6 & 12.7 &  +1.0  &
& P \\ \\
B1930+22   & 0.144 &   57.5 & 39  & 35.9 & 12.5 &  +1.6  &                        & U \\
B2334+61   & 0.495 &  191.8 & 40  & 34.8 & 13.0 &  +0.2  & X                      & P \\
J0633+1746 & 0.287 &  104.6 & 43  & 35.2 & 12.7 &  +0.5  & O,X,$\gamma$,No Radio  & U \\
B1758--23  & 0.415 &  112.9 & 58  & 34.8 & 12.8 &  --0.1 & G                      & P \\
J1105--6107& 0.063 &   15.8 & 63  & 36.4 & 12.0 &  --0.9 &
& ? \\ \\
B1727--47  & 0.829 &  163.6 & 80  & 34.1 & 13.1 &  --7.7 &                        & U \\
B0611+22   & 0.334 &   59.6 & 88  & 34.8 & 12.7 &  +2.4  &                        & ? \\
B1916+14   & 1.180 &  211.4 & 88  & 33.7 & 13.2 &  +0.9  &                        & U \\
\hline
\end{tabular}
\end{center}
\end{table}
\normalsize
\tabcolsep 6pt

\begin{table}[p]
\begin{center}
\caption{Example high frequency surveys. Note, the Parkes survey 
(\protect\cite{jlm+92}) used two recording systems in parallel, hence the
duplicate parameters. Parameters for the Nancay were kindly provided by
J-F. Lestrade}
\label{surv}
\begin{tabular}{l|lllll} \hline
                & Jodrell & Parkes   & Nancay  & PKS MB  & JOD MB \\ \hline
Nbeams          &    1    &    1     &   1     &   13    &    4   \\
$b $            & $\pm 1$ & $\pm 4$  & $\pm 3$ & $\pm 5$ & $\pm 5$\\
$l_{min}$       & $-5$    & $-90$    & $-15$   & $-90$   & $20$   \\
$l_{max}$       & $100$   & $20$     & $180$   & $20$    & $100$  \\
$T_{int}$ (min) &   10    & 2.5      &   2     &   30    &   30   \\
$T_{samp}$ (ms) &   2.0   & 1.2/0.3  &  0.06   &  0.25   &  0.25  \\
Bandwidth (MHz) &    40   & 320/80   &  150    &  288    &  64    \\
Channels (MHz)  &    5    & 5.0/1.0  &  1.6    &  3.0    &  3.0   \\
$S_{sys}$ (Jy)  &   70    & 70       &  45     &  40     &  40    \\
$S_{min}$ (mJy) &   2.0   & 1.0      & 0.4     &  0.1    &  0.2   \\
Pulsars found   &   40    &  46      & ....    & ....    & ....   \\
Accn. search    &   No    &  No      & No      & Yes     & Yes    \\ \hline
\end{tabular}
\end{center}
\end{table}

\end{document}